\documentclass[aps,prd,nofootinbib,floatfix,superscriptaddress,secnumarabic,preprintnumbers,longbibliography]{revtex4-2}
\usepackage{amsmath}
\usepackage{ulem}
\usepackage{amsfonts}
\usepackage{amssymb}
\usepackage{graphicx}
\usepackage{subcaption}
\usepackage{xcolor}
\usepackage{commath}
\usepackage{hyperref}
\usepackage{comment}
\usepackage{comment}
\usepackage[utf8]{inputenc}
\usepackage{textcomp}
\usepackage{newunicodechar}
\newunicodechar{−}{-}
\usepackage{multirow}

\newcommand{\MSbar}{{\overline{\rm MS}}}
\newcommand{\pa}{\partial}
\newcommand{\gtilde}{\frac{g^2}{16 \, \pi^2}\; }
\newcommand{\qslash}{{\not{\hspace{-0.05cm}q}}}

\newcommand{\la}{\lambda}

\newcommand{\be}{\begin{equation}}
\newcommand{\ee}{\end{equation}}
\newcommand{\bea}{\begin{eqnarray}}
\newcommand{\eea}{\end{eqnarray}}
\def\slashed{{/}\mskip-10.0mu}

\begin{document}

\title{Fine-tunings and renormalization of gluino bilinear operators in lattice SYM with Stout-Smeared links}

\author{M.~Costa}
\email[]{marios.costa@cut.ac.cy}
\affiliation{Present address: Department of Mechanical Engineering and Material Science and Engineering, Cyprus University of Technology, Limassol, CY-3036, Cyprus}
\affiliation{Rinnoco Ltd, Limassol, CY-3047, Cyprus}
\affiliation{Department of Physics, University of Cyprus, Nicosia, CY-1678, Cyprus}

\author{P.~K.~Chrysanthis}
\email{panos@rinnoco.com}
\affiliation{Rinnoco Ltd, Limassol, CY-3047, Cyprus}
\affiliation{Department of Computer Science, University of Pittsburgh, Pittsburgh, PA-15260, USA}

\author{C.~Costa}
\email{costa.c@rinnoco.com}
\affiliation{Rinnoco Ltd, Limassol, CY-3047, Cyprus}

\author{S.~Dimou}
\email{salomi@rinnoco.com}
\affiliation{Rinnoco Ltd, Limassol, CY-3047, Cyprus}

\author{G.~Spanoudes}
\email[]{spanoudes.gregoris@ucy.ac.cy}
\affiliation{Department of Physics, University of Cyprus, Nicosia, CY-1678, Cyprus}

\author{H.~Panagopoulos}
\email[]{panagopoulos.haris@ucy.ac.cy}
\affiliation{Department of Physics, University of Cyprus, Nicosia, CY-1678, Cyprus}

\begin{abstract}
In this paper, we compute the renormalization factors for the gluino and gluon fields, the gauge parameter, the coupling constant, as well as the scalar, pseudoscalar, and axial-vector gluino bilinear operators in ${\cal{N}} =1$ supersymmetric Yang–Mills (SYM) theory, using improved lattice actions. Our lattice formulation employs clover fermions, a Symanzik-improved gauge action, and stout-smeared links, which suppress ultraviolet fluctuations and thus enable more accurate determinations of renormalization factors. Our methodology involves computing gauge-variant two-point and three-point Green’s functions at one-loop order in lattice perturbation theory, in order to extract the multiplicative renormalization factors and the critical gluino mass. By analyzing lattice discretization effects on the axial current and their dependence on the stout-smearing and clover parameters, we identify a value of the smearing parameter that ensures axial-current conservation at one loop. The results presented in this work provide practical guidance for the fine-tuning procedures required to set up and calibrate lattice simulations of SYM.
\end{abstract}

\maketitle

\section{Introduction}
\label{intro}

Lattice formulations of Gauge Theories have become indispensable tools for nonperturbative studies of strong interactions, offering a robust framework for investigating gauge theories beyond the reach of perturbation theory~\cite{Wilson:1974sk, Montvay:1994cy}. These approaches allow for first-principles calculations of fundamental properties of quantum fields, such as hadronic masses, decay constants, and the running of coupling constants. Discretizing Quantum Field Theories (QFTs) on a finite space-time lattice,  circumvents the limitations of perturbative expansions and provides a controlled environment for exploring the nonperturbative regime of quantum interactions. In the same spirit, we employ Symanzik improved actions for gluons.

One of the primary challenges in lattice QFTs is the presence of discretization artifacts, which arise due to the finite lattice spacing. To mitigate these effects and enhance numerical stability, various smearing techniques have been developed. Among them, stout smearing~\cite{Morningstar:2003gk} has emerged as a particularly effective method for reducing lattice artifacts. By systematically suppressing ultraviolet fluctuations, stout smearing improves the accuracy and reliability of lattice calculations, especially in renormalization studies; this is achieved by modifying the gauge links in a controlled manner, smoothing out short-distance fluctuations while preserving gauge invariance.

Formulating supersymmetry (SUSY) on the lattice presents additional challenges, particularly in preserving both SUSY and translation invariance~\cite{Giedt:2006pd, Bergner:2009vg}. In the continuum, SUSY ensures a delicate balance between bosonic and fermionic degrees of freedom that cancels quadratic divergences in quantum corrections~\cite{Wess:1992cp}. However, discretization explicitly breaks SUSY, necessitating fine-tuning to restore it in the continuum limit. The inclusion of Majorana fermions in the adjoint representation of the gauge group (gluinos) introduces further complications, as lattice artifacts can distort these SUSY relations. Addressing these challenges requires careful design of the discretization scheme, precise renormalization of fields and operators, and appropriate parameter tuning to recover the desired continuum theory. It is also important to examine the emergence of new gauge-invariant composite operators built from a single gluino field coupled to gluon fields and to gluino bilinears. In particular, the calculation of Green’s functions of gluino bilinear operators plays a central role in understanding the nonperturbative dynamics and symmetry structure of SYM~\cite{Ali:2018dnd, Bergner:2013nwa}. The study of these operators offers insight into supersymmetry-breaking mechanisms and their potential implications for physics beyond the Standard Model~\cite{ParticleDataGroup:2024cfk}. Notably, gluino fields and their composite operators have been proposed as dark-matter candidates, making their investigation relevant to both astrophysical observations and cosmological models~\cite{Baer:1998pg}.

In this work, we focus on the lattice formulation of ${\cal N}=1$ supersymmetric Yang–Mills (SYM) with stout-smeared links and systematically analyze the required fine-tuning via lattice perturbation theory~\cite{Politzer:1973fx, Alles:1993dn, Capitani:2002mp, Skouroupathis:2008mf, Monahan:2013dla, Constantinou:2013ada, Constantinou:2013pba, Costa:2018mvb}. In our perturbative computations, the stout parameter is kept arbitrary, allowing our results to serve as a general starting point for numerical simulations, where one may make specific choices for its values. We consider three standard Symanzik-improved gluon actions and report results for each. In particular, we compute the one-loop renormalization factors for the gluino and gluon fields, the gauge-fixing parameter, the gauge coupling, and various gluino bilinear operators. This is accomplished by calculating two- and three-point Green’s functions in both dimensional regularization (DR) and lattice regularization (LR). Dimensional regularization enables a consistent definition of renormalized quantities in the $\MSbar$ scheme. 

Therefore, the novelty and difficulty of this one loop lattice calculation lie in carrying out a complete perturbative fine tuning of the SYM action while incorporating stout smeared links in the covariant derivatives of the Wilson fermion action and employing Symanzik improved gluon actions. To further optimize the discretization, we also include a clover term. In our implementation the clover field strength is built from unsmeared links, which avoids excessive smearing-induced contributions to the fermion operator and thus helps keep the gluino Pfaffian closer to its continuum form in numerical simulations~\cite{Horsley:2008ap}. From the computation of appropriate two-point and three-point Green's functions we derive analytic expressions for all renormalization factors. Our results are presented in fully general form, with the number of colors $N_c$, the gauge coupling $g$, the gauge parameter $\alpha$, and the lattice parameters, namely the clover coefficient $c_{\rm SW}$ and the stout smearing parameter $\omega$, left unspecified. Results are provided for the Wilson parameter $r=\pm 1$.

The paper is organized as follows: In Section~\ref{Dis} we present an improved lattice formulation of Supersymmetric Yang-Mills theory using stout-smeared links. This section details the discretization scheme and computational setup, explicitly defining the lattice action and relevant quantities. In Section~\ref{Green} we compute the corresponding two-point and three-point Green's functions at one-loop order. More specifically, in Section~\ref{DRGreen} we present the results computed in dimensional regularization, which are essential for defining renormalized quantities in the $\MSbar$ scheme. Subsection~\ref{latticeGreen} provides our one-loop results for lattice-regularized Green's functions, illustrating the dependence of the renormalization factors on the stout-smearing parameter, the clover coefficient, with the Wilson parameter restricted to $r=\pm 1$. These expressions offer valuable insight for the fine-tuning required to recover the continuum limit. Finally, in Section~\ref{future}, we summarize our conclusions and outline future directions, with particular emphasis on the renormalization and mixing of key supersymmetric operators, such as the Noether supercurrent and ``gluino-glue" operators. We also provide an appendix (Appendix~\ref{majoranaS}) that elaborates the Majorana nature of the gluino field within the functional integral framework.

\section{Discretization and computational setup}
\label{Dis}

In this section, we describe the discretization and computational framework employed in our lattice formulation. We explicitly define the lattice action, fields, and relevant parameters necessary for the perturbative calculation of Green's functions.

In the Wess-Zumino (WZ) gauge, the SYM action involves both the gluon and gluino fields. For clarity, we briefly outline our notation. Although the continuum SYM action used in this calculation is widely available in the literature, e.g., Refs.~\cite{Wess:1992cp, Martin:1997ns, Bergner:2022wnb, Costa:2020keq}, we briefly present it here for completeness. In the continuum and in Minkowski space, the SYM action takes the following form:
\be
{\cal S}_{\rm SYM} =  \int d^4x \Big[ -\frac{1}{4}u_{\mu \nu}^{\alpha} {u^{\mu \nu}}^{\alpha} + \frac{i}{2} \bar \lambda^{\alpha}  \gamma^\mu {\cal{D}}_\mu\lambda^{\alpha} \Big],  \, \quad u_{\mu \nu} = \pa_{\mu}u_{\nu} - \pa_{\nu}u_{\mu} + i g\, [u_{\mu},u_{\nu}],
\label{susylagr}
\ee
where $u_\mu$ is the gluon field, $\lambda$ is the gluino field which is a Majorana spinor with the property:
\begin{equation}
(\bar{\lambda}^\alpha)^T = \mathcal{C}\, \lambda^\alpha,
\label{Maj}
\end{equation}
where $\mathcal{C}$ is the charge conjugation matrix, which satisfies the relations: 
\be
(\gamma^\mu)^T \mathcal{C}^T = \mathcal{C} \gamma^\mu, \quad \mathcal{C}^{\dagger}\mathcal{C}=1, \quad \mathcal{C}^T = -\mathcal{C} \quad \Rightarrow \quad \mathcal{C} \gamma^5 = (\gamma^5)^T \mathcal{C}.
\label{Cproperties}\ee
 Besides an implicit color index, $\alpha$,  gluino fields also carry a Dirac index which is understood. The definitions of the covariant derivatives and of the gluon field tensor are:
\be
{\cal{D}}_\mu \lambda =  \pa_{\mu} \lambda + i g \,[u_{\mu},\lambda] 
\ee
The above action is invariant under the supersymmetry transformations ($\xi$ is a Majorana spinor):
\bea
\delta_\xi u_\mu^{\alpha} & = & -i \bar \xi  \gamma^\mu \lambda^{\alpha} , \nonumber \\
\delta_\xi \lambda^{\alpha}  & = & \frac{1}{4} u_{\mu \nu}^{\alpha} [\gamma^{\mu},\gamma^{\nu}] \xi.
\label{susytransfDirac}
\eea
From this point on, we will switch to Euclidean space, before introducing the lattice formulation. 

As in the case with the quantization of ordinary gauge theories, additional infinities will appear upon functionally integrating over gauge orbits. The standard remedy is to introduce a gauge-fixing term in the Lagrangian, along with a compensating Faddeev-Popov ghost term. The resulting Lagrangian, though no longer manifestly gauge invariant, is still invariant under Becchi-Rouet-Stora-Tyutin (BRST) transformations. This procedure of gauge fixing guarantees that Green's functions of gauge invariant objects will be gauge independent to all orders in perturbation theory. We use the ordinary gauge fixing term and ghost contribution arising from the Faddeev-Popov gauge fixing procedure:
\begin{equation}
{\cal S}^E_{GF}= \frac{1}{\alpha}\int d^4x \, \text{Tr} \left( \partial_\mu u_\mu\right)^2,
\label{sgf}
\end{equation}
where $\alpha$ is the gauge parameter [$\alpha=1(0)$ corresponds to Feynman (Landau) gauge], and:
\begin{equation}
{\cal S}^E_{Ghost}= - 2 \int d^4x \, \text{Tr} \left( \bar{c}\, \partial_{\mu}D_\mu  c\right), 
\label{sghost}
\end{equation}
where the ghost field $c$ is a Grassmann scalar which transforms in the adjoint representation of the gauge group, and: ${\cal{D}}_\mu c =  \pa_{\mu} c + i g \,[u_\mu,c]$. This gauge fixing term breaks supersymmetry. However, given that the renormalized theory does not depend on the choice of a gauge fixing term, and given that all known regularizations, in particular the lattice regularization, violate supersymmetry at intermediate steps, one may choose this standard covariant gauge fixing term.

In Refs.~\cite{Bergner:2022wnb, Costa:2020keq}, we investigated the mixing of various composite operators under renormalization in SYM, including the Supercurrent and the gluino-glue operators, using Wilson and clover fermions. The results from the field renormalization in these works will be further utilized in the present study. Furthermore, in our previous lattice calculations for Supersymmetric QCD (SQCD)~\cite{Costa:2017rht, Costa:2018mvb, Costa:2023cqv, Costa:2024tyz}, we extended Wilson’s formulation of the QCD action to incorporate SUSY partner fields as well. In all the aforementioned works, standard discretization was employed, where gluinos are defined on the lattice sites, while gluons are represented by link variables:  
\begin{equation}
U_\mu (x) \equiv U_{x,x+\mu} = e^{i g a T^{\alpha} u_\mu^\alpha (x+a\hat{\mu}/2)}, 
\end{equation}
where $T^{\alpha}$  are the generators of the $SU(N_c)$ gauge group. The lattice spacing, $a$, has been set to unity, $a=1$, throughout; it may be reinserted explicitly whenever necessary, by dimensional analysis. This formulation explicitly breaks all SUSY generators and chiral symmetry, necessitating fine-tuning in numerical simulations. To improve chiral symmetry restoration at finite lattice spacing and to reduce the size of required fine-tunings, we now employ stout-smeared links, $\Tilde{U}_\mu$. The stout smearing procedure improves the behavior of operator matrix elements by systematically reducing ultraviolet fluctuations in the gauge fields~\cite{Morningstar:2003gk}. It is typically applied iteratively, starting from the original link variables $U_\mu$, which are modified according to the transformation~\cite{Morningstar:2003gk, Horsley:2008ap}:
\begin{equation}
\Tilde{U}_\mu(x) = e^{i\,Q_\mu(x)}\, U_\mu(x),
\end{equation}  
where $Q_\mu(x)$ is given by:
\begin{equation}
  Q_\mu(x) = \frac{\omega}{2\,i} \left[V_\mu(x) U_\mu^\dagger(x) -  
  U_\mu(x)V_\mu^\dagger(x) -\frac{1}{N_c} \text{Tr} \,\Big(V_\mu(x)  
  U_\mu^\dagger(x) -  U_\mu(x)V_\mu^\dagger(x)\Big)\right] \, .
\label{Q_def}
\end{equation}  
Here, the ``stout coefficient'', $\omega$, is a tunable parameter, while $V_\mu(x)$ represents the sum over all staples associated with the link $U_\mu(x)$\,:
\be
V_\mu(x) = \sum_{\nu\ne\mu} \bigl(U_\nu(x) U_\mu(x+\nu) U_\nu^\dagger(x+\mu) + U_\nu^\dagger(x-\nu) U_\mu(x-\nu) U_\nu(x-\nu+\mu)\bigr)
\ee

In our lattice formulation, gluinos are represented by clover-improved Wilson fermions in the adjoint representation, with stout-smeared link variables present in their covariant derivatives. For the gauge fields we employ a class~\cite{Horsley:2008ap} of Symanzik improved gauge actions, involving Wilson loops with 4 and 6 links ($1\times1$ plaquettes and $1\times2$ rectangles, respectively). The Euclidean action ${\cal S}^{LR}_{\rm SYM}$ on the lattice then takes the form:
\begin{equation}
\begin{split}
{\cal S}^{LR}_{\rm SYM} \;=\; \sum_{x} \Bigg\{ 
& \frac{2}{g^2} \Bigg[ 
   {\rm{c_0}} \sum_{\rm plaq} \, {\rm Re\,Tr}\!\left( 1 - U_{\rm plaq} \right) 
 + {\rm{c_1}} \sum_{\rm rect} \, {\rm Re\,Tr}\!\left( 1 - U_{\rm rect} \right) 
\Bigg] \\[6pt]
& + \sum_{\mu} \Bigg[
   {\rm Tr}\!\left( \bar\lambda \gamma_{\mu} \Tilde{D}_{\mu} \lambda \right)
 - \frac{ r}{2} \, {\rm Tr}\!\left( \bar\lambda \Tilde{D}^{2} \lambda \right)
\Bigg] - \sum_{\mu,\nu} \Bigg[
   \frac{c_{\rm SW}}{4} \, 
   \bar\lambda^{\alpha} \sigma_{\mu \nu} \hat{F}_{\mu \nu}^{\alpha \beta} \lambda^{\beta}
\Bigg]
\Bigg\},
\end{split}
\label{susylagrLattice}
\end{equation}
where  $\sigma_{\mu\nu}=\frac{1}{2}[\gamma_\mu,\gamma_\nu]$ and $\hat{F}_{\mu \nu}^{\alpha \beta}$ in the adjoint representation is defined as:
\bea
\hat{{F}}_{\mu \nu}^{\alpha \beta}&=&\frac{1}{8}({Q}_{\mu \nu}^{\alpha \beta}-{Q}_{\nu \mu}^{\alpha \beta}),\\
{Q}_{\mu \nu}^{\alpha \beta}&=&2{\rm{tr}}_c \bigg( 
T^\alpha\, U_{x,x+\mu}U_{x+\mu,x+\mu+\nu}U_{x+\mu+\nu,x+\nu}U_{x+\nu,x}T^\beta\,U_{x,x+\nu}U_{x+\nu,x+\mu+\nu}U_{x+\mu+\nu,x+\mu}U_{x+\mu,x} \nonumber\\
&&\phantom{{\rm{tr}}_c } + T^\alpha\, U_{x,x+\nu}U_{x+\nu,x+\nu-\mu}U_{x+\nu-\mu,x-\mu}U_{x-\mu,x} T^\beta\, U_{x,x+\mu}U_{x+\mu,x+\mu-\nu}U_{x+\mu-\nu,x-\mu}U_{x-\nu,x} \nonumber\\
&&\phantom{{\rm{tr}}_c } + T^\alpha\, U_{x,x-\mu}U_{x-\mu,x-\mu-\nu}U_{x-\mu-\nu,x-\nu}U_{x-\nu,x} T^\beta\, U_{x,x-\nu}U_{x-\nu,x-\mu-\nu}U_{x-\mu-\nu,x-\mu}U_{x-\mu,x}\nonumber\\
&&\phantom{{\rm{tr}}_c} + T^\alpha\, U_{x,x-\nu}U_{x-\nu,x-\nu+\mu}U_{x-\nu+\mu,x+\mu}U_{x+\mu,x} T^\beta\, U_{x,x-\mu}U_{x-\mu,x-\mu+\nu}U_{x-\mu+\nu,x+\nu}U_{x+\nu,x}\bigg),
\eea
The plaquette ($U_{\rm plaq}$) and rectangle ($U_{\rm rect}$) loop variables are given by:
\begin{align}
U_{\rm plaq}(x;\mu,\nu) &= U_\mu(x)\, U_\nu(x+\hat\mu)\, 
U_\mu^\dagger(x+\hat\nu)\, U_\nu^\dagger(x), \\
U_{\rm rect}(x;\mu,\nu) &= U_\mu(x)\, U_\mu(x+\hat\mu)\, 
U_\nu(x+2\hat\mu)\, U_\mu^\dagger(x+\hat\nu+\hat\mu)\,
U_\mu^\dagger(x+\hat\nu)\, U_\nu^\dagger(x),
\end{align}
the coefficients ${\rm{c_0}} $ and ${\rm{c_1}}$ can in principle be chosen arbitrarily, subject to the following normalization condition, which ensures the correct classical continuum limit of the action:
\begin{equation}
{\rm{c_0}}  + 8 \,{\rm{c_1}}   = 1
\label{norm}
\end{equation}
For the numerical evaluation, particular choices of values for $\{c_0$, $c_1\}$ are employed in our calculations as shown in Table \ref{tb:Symanzik_coeff}.
\begin{table}[h!]
\centering
\renewcommand{\arraystretch}{1.15}
\begin{tabular}{lcc}
\hline
Gluon action & ${\rm{c_0}} $ & ${\rm{c_1}}$ \\
\hline
Wilson & $1$ & $0$ \\ 
Tree-level Symanzik (TLS)$\;\;$ & $5/3$ & $-1/12$ \\ 
Iwasaki & $\;\;3.648\;\;$ & $\;\;-0.331\;\;$ \\
\hline
\end{tabular}
\caption{Commonly used sets of values for the Symanzik coefficients.}
\label{tb:Symanzik_coeff}
\end{table}

The 4-vector $x$ is restricted to the values $x = na$, with $n$ being an integer 4-vector. The terms proportional to the Wilson parameter, $r$, eliminate the problem of fermion doubling, at the expense of breaking chiral invariance\footnote{In what follows, we will set $|r|=1$.}. In the limit $a \to 0$ the classical lattice action reproduces the continuum one. A gauge-fixing term, together with the compensating ghost field term, must also be added to the action; these terms are the same as in the non-supersymmetric case. Similarly, a standard ``measure'' term must be added to the action, in order to account for the Jacobian in the change of integration variables: $U_\mu \to u_\mu$\,. All the details and definitions of the lattice actions can be found in Ref.\cite{Costa:2017rht}.

The definitions of the covariant derivatives are as follows:
\bea
 \Tilde{{\cal{D}}}_\mu\lambda(x) &\equiv& \frac{1}{2} \Big[ \Tilde{U}_\mu (x) \lambda (x +   \hat{\mu}) \Tilde{U}_\mu^\dagger (x) - \Tilde{U}_\mu^\dagger (x -  \hat{\mu}) \lambda(x -  \hat{\mu}) \Tilde{U}_\mu(x -  \hat{\mu}) \Big], \\
 \Tilde{{\cal D}}^2 \lambda(x) &\equiv&  \sum_\mu \Big[ \Tilde{U}_\mu (x)  \lambda (x + \hat{\mu}) \Tilde{U}_\mu^\dagger (x)  - 2 \lambda(x) +  \Tilde{U}_\mu^\dagger (x -  \hat{\mu}) \lambda (x -  \hat{\mu}) \Tilde{U}_\mu(x -  \hat{\mu})\Big].
\eea
For completeness, we also present our conventions for Fourier transformations: 
\be
u_\mu(x) = \int \frac{d^4q}{(2\pi)^4}\, e^{i q \cdot x}\,\tilde{u}_\mu(q), \quad
\la(x) = \int \frac{d^4q}{(2\pi)^4}\, e^{i q \cdot x}\,\tilde{\la}(q), \quad
c(x) = \int \frac{d^4q}{(2\pi)^4 } e^{i q \cdot x}\,\tilde c(q)
\label{Fourier}
\ee

In this work we compute lattice-regularized, bare amputated two-point and three-point Green’s functions, evaluated to one-loop order in perturbation theory and to lowest order in the lattice spacing. These Green’s functions are necessary for extracting renormalization factors and investigating operator mixing phenomena in SYM theory. Specifically, we calculate the gluino propagator in momentum space: 
\begin{equation}
\langle \tilde{\lambda}^{\alpha}(q)\,\tilde{\bar{\lambda}}^{\beta}(q') \rangle,
\end{equation}
which allows the extraction of the gluino field renormalization factor, $Z_{\lambda}$, and the gluino critical mass, $m_{\lambda}^{\text{crit.}}$. Similarly, to determine the gluon field renormalization factor, $Z_u$, and renormalization of the gauge parameter, $Z_\alpha$, we compute the gluon propagator:
\begin{equation}
\langle \tilde{u}_\mu^{\alpha}(q)\, \tilde{u}_\nu^{\beta}(q') \rangle.
\end{equation}
The most straightforward way of obtaining the renormalization factor for the gauge coupling, $Z_g$, involves the amputated gluon-antighost-ghost Green's function: 
\begin{equation}
\langle \tilde{c}^{\alpha}(q)\,\tilde{\bar}{c}^{\beta}(0)\,\tilde{u}_\mu^{\gamma}(q') \rangle_{\text{amp}}.
\end{equation}
This calculation also requires the ghost propagator:
\begin{equation}
\langle \tilde{c}^{\alpha}(q)\,\tilde{\bar{c}}^{\beta}(q') \rangle,
\end{equation}
since the ghost field renormalization factor, $Z_c$, is necessary for consistently extracting $Z_g$ from the renormalization condition of Eq.~(\ref{GF3conditionlatt}). 

Additionally, we compute the amputated two-point Green’s functions involving gluino bilinear operators, defined as:
\begin{equation}
 \langle \tilde{\lambda}^{\alpha}(q)\,{\cal{O}}_{\Gamma}\,\tilde{\bar{\lambda}}^{\beta}(q') \rangle_{\text{amp}},   
\end{equation}
where the bilinear operator is given by:
\begin{equation}
{\cal{O}}_{\Gamma}= \int d^4x \,\text{Tr}\left(\bar{\lambda}(x)\,\Gamma\,\lambda(x) \right),
\end{equation}
with the understanding that, on the lattice, the spacetime integration is replaced by a sum over lattice points. The matrix $\Gamma$ denotes a generic element of the Clifford algebra of Dirac matrices, which in principle can be chosen from the following set:
\begin{equation}
\Gamma \in \{ \openone,\,\gamma_5,\,\gamma_\mu,\,\gamma_5\,\gamma_\mu,\,\sigma_{\mu \,\nu} \} \equiv \{S,\,P,\,V,\,A,\,T\}.
\end{equation}
However, using the Majorana condition in Eq.~(\ref{Maj}), the gluino bilinears are rewritten as:
\begin{equation}
{\cal{O}}_{\Gamma}\equiv\int d^4x \,\text{Tr}\left(\bar{\lambda}^\alpha(x) T^\alpha\,\Gamma\,\lambda(x)^\beta T^\beta \right)=  \frac{1}{2}\, 
 \int d^4x \,\bar{\lambda^\alpha}^T(x)\,\mathcal{C}^T\,\Gamma\,\lambda^\alpha(x).
\end{equation}
Because the components of $\lambda$ are Grassmann variables, exchanging the two $\lambda$'s introduces a minus sign, which implies:
\be
\lambda^T \mathcal{C}^T \Gamma \lambda = -\, \lambda^T (\mathcal{C}^T \Gamma)^T \lambda .
\ee
Therefore this expression can be nonzero only if $\mathcal{C}^T \Gamma$ is antisymmetric. Using the standard transpose identities for the Dirac basis,
one finds that $\mathcal{C}^T \openone$, $\mathcal{C}^T \gamma_5$, and $\mathcal{C}^T \gamma_5 \gamma_\mu$
are antisymmetric, while $\mathcal{C}^T \gamma_\mu$ and $\mathcal{C}^T \sigma_{\mu\nu}$ are
symmetric. Consequently, the vector $V$ and tensor $T$ currents vanish identically for a gluino bilinear, and only the scalar $S$, pseudoscalar $P$, and axial vector $A$, bilinears can be nonzero.

In general, all these two-point and three-point Green’s functions are calculated using both dimensional (continuum) and lattice regularization. The continuum Green’s functions are used to obtain the renormalized Green’s functions in the $\MSbar$ scheme, which serve as essential inputs for defining renormalization conditions on the lattice. The corresponding renormalization factors are defined as follows:
\bea
\la^R &=& \sqrt{Z_\la}\,\la^B,\\
u_{\mu}^R &=& \sqrt{Z_u}\,u^B_{\mu},\\
c^R &=& \sqrt{Z_c}\,c^B, \\
g^R &=& Z_g\,\mu^{-\epsilon}\,g^B,\\
\alpha^R &=& Z_\alpha^{-1} \, Z_u \, \alpha^B,\\
{\cal{O}}_\Gamma^R &=& Z_{{\cal{O}}_\Gamma} {\cal{O}}_\Gamma^B, \quad \Gamma \equiv \{S,\,P,\,A \},
\eea
where $B$ stands for bare and $R$ for renormalized quantities, and $\mu$ is an arbitrary scale with dimensions of inverse length. For one-loop calculations, the distinction between $g^R$ and $g^B$ is inessential in many cases; we will simply use $g$ in those cases. Although the gluino is massless in the continuum SYM action, Wilson-type lattice discretizations break chiral symmetry and generate an additive mass renormalization. This leads to a power-divergent contribution, inversely proportional to the lattice spacing, denoted by $m_\lambda^{\rm crit.}$, which must be tuned non-perturbatively in order to restore chiral symmetry and supersymmetry in the continuum limit~\cite{Curci:1986sm}. 

\section{Computation of Green's functions and Results}
\label{Green}

In this section, we present our detailed computations of the two-point and three-point Green's functions at one-loop order, discussing both dimensional ($DR$) and lattice ($LR$) regularizations, and present the resulting analytic expressions for the renormalization factors. To avoid heavy notation, we will omit the tilde from the Fourier-transformed fields. These fields will be understood from their arguments. 

\subsection{Dimensional Regularization (DR)}
\label{DRGreen}

In this subsection, we calculate the propagators, ghost vertex correction, and Green's functions for gluino bilinear operators in the continuum, regularizing the theory in $D$ Euclidean dimensions ($D=4-2\epsilon$). Several prescriptions for defining $\gamma_5$ in $D$ dimensions exist in the literature; they are related among themselves via finite conversion factors~\cite{Buras:1989xd}. In our continuum calculation, since $\MSbar$-renormalized Green’s functions are computed in Dimensional Regularization (DR), we adopt the 't Hooft--Veltman (HV) prescription~\cite{tHooft:1972tcz} for defining $\gamma_5$ in $D$ dimensions:
\be
\{\gamma_5,\gamma_{\mu}\} = 0, \, \,\mu = 1,2,3,4, \qquad [\gamma_5,\gamma_{\mu}]=0, \,\, \mu>4.
\ee
This prescription does not lead to algebraic inconsistencies with $\gamma_5$-odd
fermion traces, as in the naive prescription of DR (NDR).
The tensor metric $\delta_{\mu\nu}$, and the Dirac matrices, $\gamma_\mu$, satisfy the following relations in $D$ dimensions:
\be
\delta_{\mu\nu}\delta_{\mu\nu}=D,\qquad \{\gamma_\mu,\gamma_\nu\} = 2 \delta_{\mu\nu} \openone.
\ee

We first present the inverse gluino propagator, whose quantum corrections are computed from the first diagram shown in Fig.~\ref{gluinoProp}: 
\begin{equation}
\langle \lambda^{\alpha} (q)\, \bar \lambda^{\beta}(q') \rangle^{B, DR}_{\rm{inv}} =  (2\pi)^4 \delta(q-q') \frac{i}{2} \delta^{\alpha \beta} \frac{\qslash}{2} \left[ 1 + \frac{g^2 N_c}{16 \pi^2} \alpha   \left(\frac{1}{\epsilon} + 1 + \log\left(\frac{\bar \mu^2}{q^2} \right)\right)\right],
\label{continuumGluino}
\end{equation}
where $N_c$ is the number of colors, $q$ is the external momentum in the Feynman diagrams, and $\bar\mu$ is the energy scale which is related to $\mu$ through\footnote{$\gamma_E$  is Euler's constant: $\gamma_E = 0.57721\ldots$}: $\mu = \bar \mu \sqrt{e^{\gamma_E}/ 4\pi}$.

We now turn to the inverse gluon propagator. Individually, the continuum diagrams in the first row of Fig.~\ref{gluonProp} yield non-transverse contributions. However, their sum restores transversality, resulting in the following continuum expression:
\bea
\label{GF2gluoncont}
\langle  u_\mu^{\alpha}(q) u_\nu^{\beta}(q') \rangle^{B, DR}_{\rm{inv}} &=&(2\pi)^4 \delta(q+q') \delta^{\alpha \beta}\Bigg\{  \frac{1}{\alpha} q_{\mu} q_{\nu} + \left(q^2 \delta_{\mu \nu} - q_{\mu} q_{\nu}\right) \times \\ \nonumber
&& \Bigg[ 1 - \frac{g^2\, N_c}{16\,\pi^2}\frac{1}{2} \left(\left(3-\alpha\right)\frac{1}{\epsilon} + \frac{19}{6} + \alpha + \frac{\alpha^2}{2} + \left(3-\alpha\right)\log\left(\frac{\bar\mu^2}{q^2} \right)\right)\Bigg] \Bigg\}
\eea
Since there is no one-loop longitudinal part for the gluon self-energy, the renormalization factor for the gauge parameter receives no one-loop contribution. 

Finally, we present the inverse ghost propagator, shown in Fig.~\ref{ghostProp}, which is identical to its non-supersymmetric counterpart since it receives contributions exclusively from the ghost action $S_{Ghost}$, involving ghost and gluons fields:
\be
\langle c^{\alpha}(q) {\bar{c}}^{\beta}(q') \rangle^{B, DR}_{\rm{inv}} = (2\pi)^4 \delta(q-q') \delta^{\alpha \beta}  q^2 \left[ 1 - \frac{g^2\,N_c}{16\,\pi^2} \left(1 + \frac{3-\alpha}{4\epsilon}  + \frac{1}{4} (3-\alpha) \log\left(\frac{\bar\mu^2}{q^2} \right) \right) \right].
\label{GF2ghost}
\ee

Let us now calculate the amputated gluon-antighost-ghost Green's function in order to renormalize the coupling constant. Other determinations of the coupling constant renormalization (e.g., through gluon-gluino-gluino and three-gluon Green's functions) are expected to lead to identical results. We compute, perturbatively to one-loop, this Green's function using both dimensional and lattice regularizations. In the first row of Fig.~\ref{gluonghostantighost}, we have drawn the corresponding continuum 1PI (one-particle irreducible) Feynman diagrams. The three-point amputated Green's function, at zero antighost momentum, in $DR$ gives:
\be
\langle c^{\alpha}(q) {\bar{c}}^{\beta}(0) {u}_\mu^{\gamma}(q')\rangle^{B, DR}_{\rm{amp}}  = 
(2\pi)^4 \delta(q+q') f^{\alpha\,\beta\,\gamma} (i g q_\mu) \left[1 + \frac{g^2\,N_c}{16\,\pi^2}\frac{\alpha}{2} \left( 1 + \frac{1}{\epsilon}  +  \log \left(\frac{\bar{\mu}^2}{q^2} \right) \right) \right].
\label{GF3ghostsgluon}
\ee

A primary aim of this work is the computation of the renormalization of bilinear composite operators constructed from gluino fields. These operators are not only key to understanding the dynamics of supersymmetric gauge theories, but they are also considered strong candidates for describing dark matter particles~\cite{Flores:1990bt, Falkowski:2009yz, Catena:2013pka, Acharya:2017szw, Cacciapaglia:2023syp}.

\begin{figure}[ht!]
\centering
\includegraphics[scale=0.32]{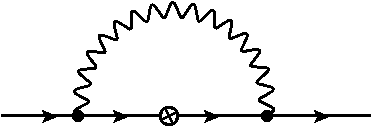} 
\caption{One-loop Feynman diagram contributing to the two-point Green's functions of the gluino bilinear operators $\langle \la \, {\cal{O}}_\Gamma \, \bar \la  \rangle$\,. A wavy (solid) line represents gluons (gluinos). A cross denotes the insertion of ${\cal{O}}_\Gamma$.}
\label{bilinear}
\end{figure}

The corresponding bare two-point Green's functions of gluino bilinears are first evaluated using dimensional regularization via the Feynman diagram shown in Fig.~\ref{bilinear}. The Green's functions of the three bilinears give:
\bea
\label{scalarB}
\langle \la^{\alpha}(q) \, {\cal{O}}_S \, \bar \la^{\beta}(q')\rangle^{B, DR}_{\rm{amp}} &=& (2\pi)^4 \delta(q - q')\, \frac{1}{4} \delta^{\alpha \beta}
\left[\openone + \openone \frac{g^2 N_c}{16 \pi^2} \left( \frac{3 + \alpha}{\epsilon} + 4 + 2 \alpha + (3 + \alpha) \log \left(\frac{\bar\mu^2}{q^2} \right) \right) \right], \\[1ex]
\label{pseudoscalarB}
\langle \la^{\alpha}(q) \, {\cal{O}}_P \, \bar \la^{\beta}(q')\rangle^{B, DR}_{\rm{amp}} &=& (2\pi)^4 \delta(q - q')\, \frac{1}{4} \delta^{\alpha \beta}
\left[ \gamma_5 + \gamma_5  \frac{g^2 N_c}{16 \pi^2} \left(\frac{3 + \alpha}{\epsilon} + 12 +2  \alpha  + (3 + \alpha) \log \left(\frac{\bar\mu^2}{q^2} \right)\right) \right],\\[1ex]
\label{axialvectorB}
\langle \la^{\alpha}(q) \, {\cal{O}}_A \, \bar \la^{\beta}(q')\rangle^{B, DR}_{\rm{amp}} &=& (2\pi)^4 \delta(q - q')\, \frac{1}{4} \delta^{\alpha \beta}
\left[ \gamma_5 \gamma_\mu + \frac{g^2 N_c}{16 \pi^2} \left( \gamma_5 \gamma_\mu  \left( \frac{\alpha}{\epsilon} + 4 + \alpha + \alpha \log \left(\frac{\bar\mu^2}{q^2} \right) \right)\right) \nonumber \right. \\
&&\left. \phantom{\left( \gamma_5 \gamma_\mu  \left( \frac{\alpha}{\epsilon} + 4 + \alpha + \alpha \log \left(\frac{\bar\mu^2}{q^2} \right) \right)\right)}  -\frac{g^2 N_c}{16 \pi^2} 2 \alpha \gamma_5 \frac{q_\mu \slashed{q}}{q^2}  \right].
\label{GFBilinearSimplified}
\eea

The continuum Green's functions computed in DR provide the reference amplitudes needed to define renormalized quantities in the $\MSbar$ scheme. They serve as benchmark results for setting lattice renormalization conditions and for matching lattice regularized operators and parameters to their continuum counterparts. This ensures consistency between the two regularizations.

At one-loop order, the two-point Green's function involving the axial-vector current in Eq.~(\ref{axialvectorB}) develops an additional tensor structure beyond its tree level Dirac form. A similar nonmultiplicative term also appears in the corresponding lattice Green's function. While this current is classically conserved, quantum effects, in particular the axial anomaly, modify the Ward identities and lead to the appearance of this extra structure even in the massless limit.

In DR and in the $\MSbar$ scheme, the determination of field and operator renormalization factors $Z^{DR,\MSbar}$ is such that $\MSbar$ renormalized Green's functions, obtained by multiplying the corresponding bare Green's functions by the appropriate renormalization factors, must coincide with the bare Green's functions after subtracting the $1/\epsilon$ poles from the latter and taking the limit $\epsilon \to 0$. Use of the $\MSbar$ renormalized Green's functions will be made in the left-hand sides of Eqs.~(\ref{GG2gluinocondition}), (\ref{GG2gluoncondition}), (\ref{GG2ghostcondition}) and (\ref{GF2Bilinearconf}), in order to define the lattice renormalization factors $Z^{LR,\MSbar}$.  An analogous rationale applies to the renormalization condition in Eq.~(\ref{GF3conditionlatt}). We thus obtain the following one loop renormalization factors:
\bea
Z_\la^{DR,\MSbar} &=& 1 + \frac{g^2\,N_c}{16\,\pi^2} \frac{1}{\epsilon} \alpha , \\
Z_u^{DR,\MSbar} &=&  1 + \frac{g^2\,N_c}{16\,\pi^2} \frac{1}{\epsilon} \frac{\alpha-3}{2},\\
Z_c^{DR,\MSbar} &=&  1 + \frac{g^2\,N_c}{16\,\pi^2} \frac{1}{\epsilon} \frac{\alpha-3}{4},\\
Z_g^{DR,\MSbar} &=&  1 + \frac{g^2\,N_c}{16\,\pi^2} \frac{1}{\epsilon} \frac{3}{2}, \\
Z_S^{DR,\MSbar} &=&  1 + \frac{g^2\,N_c}{16\,\pi^2} \frac{1}{\epsilon} (-3),\\
Z_P^{DR,\MSbar} &=&  1 + \frac{g^2\,N_c}{16\,\pi^2} \frac{1}{\epsilon} (-3),\\
Z_A^{DR,\MSbar} &=&  1 .
\eea
$Z_A^{DR,\MSbar}$ does not receive one-loop corrections, as it corresponds to a conserved current of the theory. However, on the lattice, this operator receives finite corrections due to the breaking of chiral and rotational symmetries by the lattice regularization. Consequently, its Green's function requires additional finite renormalization to restore the correct continuum limit.

\subsection{Lattice Regularization (LR)}
\label{latticeGreen}
We now turn to the LR framework. The lattice formulation introduces discretization effects that modify the continuum expressions, and these modifications explicitly depend on lattice parameters such as the clover coefficient $c_{\rm SW}$, the Wilson parameter $r$, the stout-smearing parameter $\omega$, and lattice spacing $a$. 

\begin{figure}[ht!]
\centering
\includegraphics[scale=0.32]{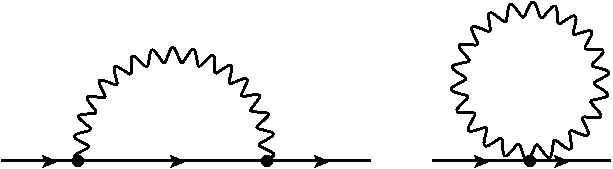} 
\caption{One-loop 1PI Feynman diagrams leading to the gluino self-energy on the lattice.  A wavy (solid) line represents gluons (gluinos). Only the first diagram contributes in  dimensional regularization. }
\label{gluinoProp}
\end{figure}

The lattice expression for the gluino inverse propagator is obtained by evaluating the one-loop Feynman diagrams illustrated in Fig.~\ref{gluinoProp}. These diagrams represent all contributing terms at this perturbative order, incorporating gluon and gluino interactions with vertices defined by the lattice action of Eq.~(\ref{susylagrLattice}). 

The resulting expression for the gluino inverse propagator is given by\footnote{The numerical errors in our one-loop lattice expressions are less than the last digit presented.}: 

\begin{align}
\label{GF2gluinolatt}
\big\langle \lambda^{\alpha}(q)\,\bar{\lambda}^{\beta}(q') \big\rangle^{B, LR}_{\rm inv}
&= (2\pi)^4 \delta(q-q') \frac{\delta^{\alpha\beta}}{2} \Bigg\{
i\,\frac{\qslash}{2}
\Bigg[ 1 - \frac{g^2 N_c}{16\pi^2}
\Big( {\rm{c}}^\la_{00} - 4.7920\,\alpha
- {\rm{c}}^\la_{10}\,c_{\rm SW} r
- {\rm{c}}^\la_{20}\,c_{\rm SW}^2 \nonumber \\
&\qquad\qquad\qquad\qquad\hspace{3cm}
- {\rm{c}}^\la_{01}\,\omega
+ {\rm{c}}^\la_{11}\,c_{\rm SW} r \,\omega
+ {\rm{c}}^\la_{02}\,\omega^2
+ \alpha \log(a^2 q^2) \Big) \Bigg] \nonumber \\
&\qquad\hspace{2.7cm}
+ \frac{\openone}{2\,a}\,  \frac{g^2 N_c}{16\pi^2}\, r \,
\Big( {\rm{c}}^m_{00}
- {\rm{c}}^m_{10}\, c_{\rm SW} r 
- {\rm{c}}^m_{20}\, c_{\rm SW}^2 \nonumber \\
&\qquad\qquad\hspace{4.2cm}
- {\rm{c}}^m_{01}\,\omega
+ {\rm{c}}^m_{11}\, c_{\rm SW}r\, \omega
+ {\rm{c}}^m_{02}\,\omega^2
\Big)
\Bigg\} + \mathcal{O}(a),
\end{align}

\begin{table}[h!]
\centering
\renewcommand{\arraystretch}{1.15}
\begin{tabular}{lcccccccccccc}
\hline
Gluon action & ${\rm{c}}^\la_{00}$ & ${\rm{c}}^\la_{10}$ & ${\rm{c}}^\la_{20}$ &  ${\rm{c}}^\la_{01}$ & ${\rm{c}}^\la_{11}$ & ${\rm{c}}^\la_{02}$ & ${\rm{c}}^m_{00}$ & ${\rm{c}}^m_{10}$ & ${\rm{c}}^m_{20}$ & ${\rm{c}}^m_{01}$ & ${\rm{c}}^m_{11}$ & ${\rm{c}}^m_{02}$ \\
\hline
Wilson   & 16.6444 & 4.4977 & 5.5891  & 202.0079 & 11.1238 & 739.6836 & 51.4347 & 27.4663 & 22.8606 & 612.3029 & 91.4422  & 2335.4071 \\
TLS      & 13.0233 & 4.0309 & 4.9688  & 152.5642 & 9.3469  & 541.3805 & 40.4432 & 23.8964 & 18.6507 & 455.5139 & 74.6028  & 1685.5974 \\
Iwasaki  & 8.1166 & 3.2020 & 3.8928  & 91.1438  & 6.5475  & 308.7532 & 26.0729 & 18.0307 & 12.4245 & 266.8277 & 49.6981  & 945.1217  \\
\hline
\end{tabular}
\caption{Numerical values of the coefficients  ${\rm{c}}^\la_{00},\,{\rm{c}}^\la_{10},\,{\rm{c}}^\la_{20},\,{\rm{c}}^\la_{01},\,{\rm{c}}^\la_{11},\,{\rm{c}}^\la_{02}$ and ${\rm{c}}^m_{00},\,{\rm{c}}^m_{10},\,{\rm{c}}^m_{20},\,{\rm{c}}^m_{01},\,{\rm{c}}^m_{11},\,{\rm{c}}^m_{02}$ for the three actions, cf. Eq.~(\ref{GF2gluinolatt}). Note that at one loop $\alpha$ enters only through the longitudinal gluon propagator, while stout smearing affects only the transverse part, so no $\alpha\times\omega$ term appears.}
\label{tb:coeff}
\end{table}

To renormalize the gluino field, we impose the condition that the renormalized propagator in the $\overline{\text{MS}}$ scheme satisfies:
\begin{equation}
\langle\lambda\,{\bar{\lambda}} \rangle ^{\MSbar}_{\rm inv} = \left(Z_{\lambda}^{LR,\MSbar}\right)^{-1}\, \langle\lambda\,{\bar{\lambda}} \rangle^{B, LR}_{\rm inv}\Big{|}_{a \to 0},
\label{GG2gluinocondition}
\end{equation}
where we use Eq.~(\ref{continuumGluino}) without the $1/\epsilon$ term to define the $\overline{\text{MS}}$ renormalized Green's function. This leads to the renormalization factor of the gluino field:

\begin{align}
Z_{\lambda}^{LR,\MSbar}
&= 1 - \frac{g^2 N_c}{16\pi^2}\Bigg[
  {\rm{c}}^\la_{00} - 3.7920\,\alpha - {\rm{c}}^\la_{10}\,c_{\rm SW}\,r - {\rm{c}}^\la_{20}\,c_{\rm SW}^2 
  + \alpha \log\!\left(a^2 \bar\mu^2\right)
  - \omega\Big( {\rm{c}}^\la_{01} - {\rm{c}}^\la_{11}\,c_{\rm SW}\,r - {\rm{c}}^\la_{02}\,\omega \Big)
\Bigg].
\end{align}

Finally, the critical mass, arising from the explicit breaking of chiral symmetry for the gluino field for the three choices of the gluon action, is given by:
\begin{align}
m_\lambda^{\rm crit.}
&= \frac{g^2 N_c}{16\pi^2}\,\frac{1}{a}\,r\,
\Big(
  {\rm{c}}^m_{00}
- {\rm{c}}^m_{10}\, c_{\rm SW} r
- {\rm{c}}^m_{20}\, c_{\rm SW}^2
+ {\rm{c}}^m_{11}\, c_{\rm SW} r\, \omega
- {\rm{c}}^m_{01}\, \omega
+ {\rm{c}}^m_{02}\, \omega^2
\Big).
\label{mcrit}\end{align}

These results highlight the lattice artifacts introduced in the discretization of the theory and provide the necessary counterterms to recover the correct continuum limit. The dependence of $m_\lambda^{\rm crit.}$ on the parameters $c_{\rm SW}$, $r=\pm1$, and $\omega$ suggests that tuning these coefficients appropriately can mitigate discretization effects and restore supersymmetry in the continuum limit~\cite{Curci:1986sm}. While for quantities exhibiting a  power divergence with respect to the lattice spacing, such as  $m_\lambda^{\rm crit.}$, a nonperturbative evaluation is essential, Eq.~(\ref{mcrit}) provides a useful ball-park estimate.

\bigskip

\begin{figure}[ht!]
\centering
\includegraphics[scale=0.32]{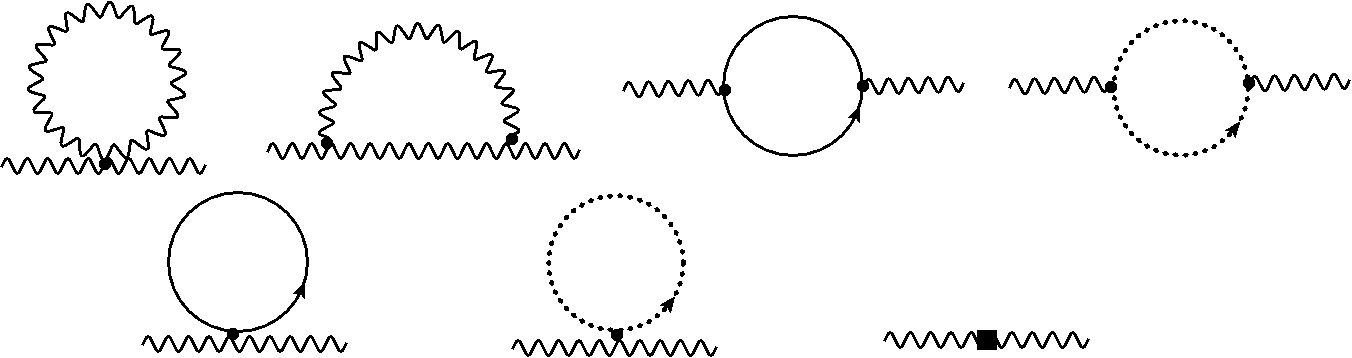} 
\caption{One-loop 1PI Feynman diagrams contributing to the gluon propagator at one-loop order. A wavy (solid, dotted) line represents gluons (gluinos, ghosts). The square vertex denotes contributions arising from measure term. The diagrams shown in the second row appear exclusively due to the lattice discretization and have no continuum counterparts.}
\label{gluonProp}
\end{figure}

The next calculation concerns the gluon propagator. The corresponding one-loop Feynman diagrams are shown in Fig.~\ref{gluonProp}. These diagrams depict contributions from gluons, gluinos, ghost fields, and the measure term, all of which must be taken into account to properly evaluate lattice discretization effects. 

The gluon inverse propagator is given to one loop by:
\bea
\label{GF2gluonlatt}
\Big\langle  u_\mu^{\alpha}(q)\, u_\nu^{\beta}(q') \Big\rangle^{B,LR}_{\rm inv}
&=& (2\pi)^4 \delta(q+q')\,\delta^{\alpha\beta} \Bigg\{
\frac{1}{\alpha}\,q_\mu q_\nu + \left(q^2 \delta_{\mu\nu} - q_\mu q_\nu\right)
\Bigg[\,1 - \frac{g^2}{16\pi^2}\Bigg(
{\rm{c}}_{-1}^{u}\,\frac{1}{N_c} \nonumber\\
&&\quad 
+ N_c\Big({\rm{c}}_1^{u}
- 0.8863\,\alpha + \frac{\alpha^2}{4}
- 18.8508\,c_{\rm SW}^2 + 1.5939\,c_{\rm SW}\,r \nonumber \\
&& \hspace{5cm}+ \Big(\frac{\alpha}{2}-\frac{3}{2}\Big)\log(a^2 q^2)
\Big)\Bigg)\Bigg]\Bigg\} + \mathcal{O}(a).
\eea

\begin{table}[h!]
\centering
\renewcommand{\arraystretch}{1.15}
\begin{tabular}{lcc}
\hline
Gluon action & ${\rm{c}}^u_{-1}$ & ${\rm{c}}^u_{1}$ \\
\hline
Wilson   & -19.7392 & +20.1472 \\
TLS      & -6.6595  & +8.1403  \\
Iwasaki  & +11.8888 & -9.4149  \\
\hline
\end{tabular}
\caption{Numerical values of the coefficients ${\rm{c}}^u_{-1}$ and ${\rm{c}}^u_{1}$ for the three gluon actions.}
\label{tb:coeff_u}
\end{table}

For the two-point Green's function $\langle  u_\mu(q) \ u_\nu(q') \rangle^{B, LR}_{\rm{inv}}$, some individual Feynman diagrams contribute a quadratically divergent mass term, proportional to $1/a^2$. However, upon summing all diagrams, these divergences cancel exactly. Another noteworthy cancellation concerns Lorentz-noncovariant terms of the form $(\delta_{\mu \nu} q^2_{\nu})$; after summing all contributions, such terms also cancel, leaving a transverse expression for the gluon self-energy. This result reflects the gauge invariance of the theory. Since the gluon self-energy contains neither a critical mass term nor a longitudinal component, the renormalization factor of the gauge parameter receives no contribution at one loop: 
\be 
Z_\alpha ^{LR,\MSbar} = 1 + {\cal{O}}(g^4). 
\ee
At one loop, $\langle  u_\mu(q)\, u_\nu(q') \rangle^{B,LR}_{\rm inv}$ is independent of the stout parameter $\omega$ to the order considered here. The same statement holds for any number of stout smearing steps. Indeed, after $N$ successive smearing steps, the transverse part of the gluon field entering one loop corrections is multiplied by a factor $(1-\omega\,\hat q^{\,2})^{N}$~\cite{Constantinou:2022aij}, where $\hat q^{\,2}=\sum_\mu \hat q_\mu^{\,2}$ and $\hat q_\mu=(2/a)\sin(a q_\mu/2)$. Since $(1-\omega\,\hat q^{\,2})^{N}=1+\mathcal{O}(a^{2})$, the $\omega$ dependence drops out when keeping only the leading terms in the lattice spacing. 

Noting also that the maximum value of $\sum_\mu (2 / a)^2 \sin^2(a p_\mu / 2)$ is 16, with an average value around 8, one can infer that choosing $\omega$ in the range $ 1/16 \lesssim \omega \lesssim 1/8 $ effectively suppresses the transverse components of the gluon fields after several smearing steps. Further, at one loop, the pure-gauge part of the gluon self-energy decomposes into two independent color structures, proportional to $1/N_c$ and $N_c$. The coefficients of the logarithms and the $c_{\rm SW}^2$ and $c_{\rm SW}\,r$ terms arise from fermion loops are independent of the gluon action; diagrams with closed ghost loops are also independent of the choice of gluon action. Different gluon actions only modify the finite pieces multiplying those two color structures, captured by ${\rm{c}}^u_{-1}$ and ${\rm{c}}^u_{1}$.

By demanding the following:
\be
\langle u_\mu\,u_\nu \rangle^\MSbar _{\rm inv} = \left(Z_{u}^{LR,\MSbar}\right)^{-1}\, \langle u_\mu \,u_\nu \rangle^{B, LR}_{\rm inv}\Big{|}_{a \to 0},
\label{GG2gluoncondition}
\ee
and using Eqs.~(\ref{GF2gluonlatt}), (\ref{GF2gluoncont}) we find:
\be
Z_{u}^{LR,\MSbar}
= 1 + \frac{g^2}{16\pi^2}
\left[
-\frac{{\rm{c}}_{-1}^{u}}{N_c}
- N_c\!\left(
{\rm{c}}_1^{u} - \frac{19}{12}
- 1.3863\,\alpha
- 18.8508\,c_{\rm SW}^2
+ 1.5939\,c_{\rm SW}\,r
+ \left( -\frac{3}{2}+\frac{\alpha}{2}\right) \log\!\left(a^2\bar\mu^2\right)
\right)
\right].
\ee

\begin{figure}[ht!]
\centering
\includegraphics[scale=0.32]{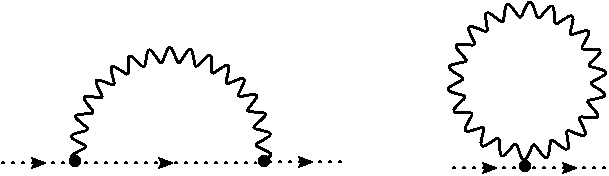} 
\caption{One-loop 1PI Feynman diagrams leading to the ghost propagator.  A wavy (solid, dotted) line represents gluons (gluinos, ghosts). The second diagram appears only in the lattice regularization. }
\label{ghostProp}
\end{figure}
Since we choose to utilize the vertex gluon-antighost-ghost for the renormalization of the coupling constant $g$, we need also the ghost field renormalization, $Z_c$, which can be extracted from the ghost propagator; calculating the Feynman diagrams shown in Fig.~\ref{ghostProp}, we find:
\be
\langle c^{\alpha}(q) {\bar{c}}^{\beta}(q') \rangle^{B, LR}_{\rm{inv}}  =  (2\pi)^4 \delta(q-q') q^2 \delta^{\alpha \beta}\left[ 1-\frac{g^2\,N_c}{16\,\pi^2}\left({\rm{c}}^c_{1} - 1.2029 \, \alpha -\frac{1}{4}\left( 3 - \alpha \right)
\log\left(a^2\,q^2\right)\right)\right]+{\cal{O}}(a),
\label{GF2ghostlatt}
\ee
and the renormalization condition reads:
\begin{equation}
\langle c\,{\bar{c}} \rangle ^{\MSbar}_{\rm inv} = \left(Z_{c}^{LR,\MSbar}\right)^{-1}\, \langle c\,{\bar{c}} \rangle^{B, LR}_{\rm inv}\Big{|}_{a \to 0},
\label{GG2ghostcondition}
\end{equation}
which gives the following value for $Z_c^{LR,\MSbar}$:
\be
Z_c^{LR,\MSbar} = 1 - \frac{g^2 N_c}{16\pi^2} \Bigl[{\rm{c}}^c_{1} - 1 - 1.2029 \, \alpha -\frac{1}{4}\left( 3 - \alpha \right)
\log\left(a^2\,\bar{\mu}^2\right) \Bigr].
\ee

\begin{table}[h!]
\centering
\renewcommand{\arraystretch}{1.15}
\begin{tabular}{lc}
\hline
Gluon action & ${\rm{c}}^c_{1}$ \\
\hline
Wilson   & 4.6086 \\
TLS      & 3.7759 \\
Iwasaki  & 2.5469 \\
\hline
\end{tabular}
\caption{Numerical values of the coefficient ${\rm{c}}^c_{1}$ for the three gluon actions.}
\label{tb:coeff_c}
\end{table}

\begin{figure}[ht!]
\centering
\includegraphics[scale=0.32]{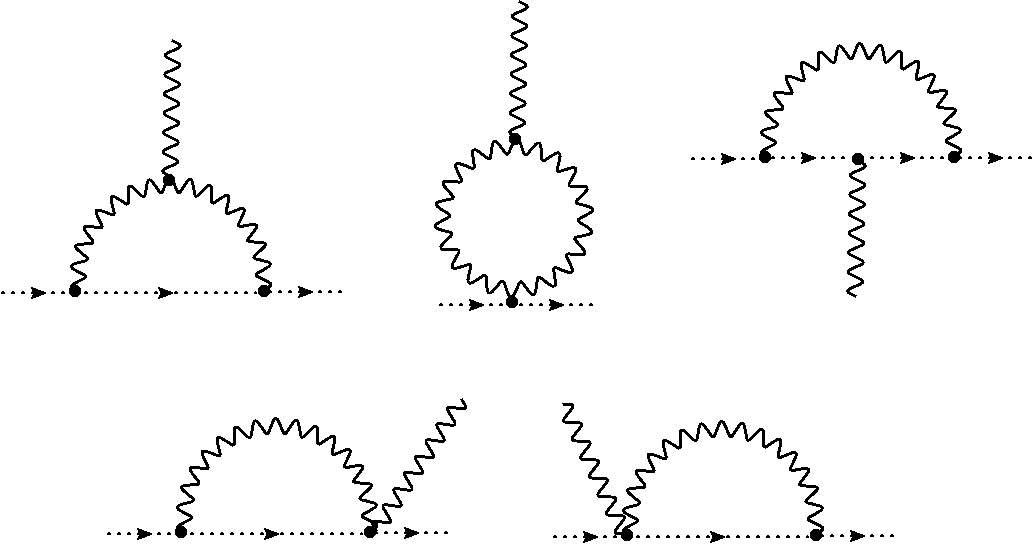} 
\caption{One-loop 1PI Feynman diagrams contributing to the renormalization of the coupling constant. Wavy lines represent gluons, and dotted lines denote ghost fields. The second diagram in the first row and the diagrams in the second row contribute exclusively in the lattice regularization.}
\label{gluonghostantighost}
\end{figure}

The Feynman diagrams for the gluon-antighost-ghost Green's function, $\langle u_\mu^{\alpha} c^{\beta} \bar c^{\gamma} \rangle^{B, LR}_{\text{amp}}$  are shown in Fig.~\ref{gluonghostantighost}. In the $\MSbar$ scheme, the renormalization condition is defined as follows:
\be
\lim_{a \to 0}\Big[\left(Z_{c}^{LR,\MSbar}\right)^{-1} \left(Z_{u}^{LR,\MSbar}\right)^{-1/2}\langle c^{\alpha}(q) \, {\bar{c}}^{\beta}(0) \,{u}_\mu^{\gamma}(q')\rangle^{B, LR}_{\rm{amp}}\Big{|}_{g \to \left(Z_{g}^{LR,\MSbar}\right)^{-1} g^R} \Big] = \langle  c^{\alpha}(q) \, {\bar{c}}^{\beta}(0) \, u_\mu^{\gamma}(q')\rangle^{\MSbar}_{\rm{amp}}
\label{GF3conditionlatt}
\ee
where the right hand side in the above equation equals $\langle c^{\alpha}(q) \,{\bar{c}}^{\beta}(0) \, u_\mu^{\gamma}(q')\rangle^{B, DR}_{\rm{amp}}\Big{|}_{1/\epsilon \to 0}$ of Eq.~(\ref{GF3ghostsgluon}), and the corresponding expression on the lattice is:
\be
\langle c^{\alpha}(q) \, {\bar{c}}^{\beta}(0) \, {u}_\mu^{\gamma}(q')\rangle^{B, LR}_{\rm{amp}}  = 
(2\pi)^4 \delta(q+q') f^{\alpha\,\beta\,\gamma} \left(i g \, q_\mu\right) \Bigg[1 + \frac{g^2\,N_c}{16\,\pi^2}\left(
2.3960 \alpha - \frac{1}{2} \alpha \log\left(a^2\, q^2 \right)\right)
\Bigg]+{\cal{O}}(a),
\ee
which holds for any choice of gluon action.

Our result for $Z_g^{LR,\MSbar}$ is:
\be
Z_g^{LR,\MSbar} =  1 + \gtilde\,\Bigg[\frac{{\rm{c}}_{-1}^{u}}{2} \frac{1}{N_c} + N_c \left(\frac{1}{2} \left({\rm{c}}_1^{u} - \frac{19}{12} \right) + {\rm{c}}_1^{c} - 1 - 9.4225 \,c_{\rm SW}^2  + 0.7969 \, c_{\rm SW}\, r - \frac{3}{2} \log\left(a^2\,\bar{\mu}^2\right)\right)\Bigg].
\ee
At one-loop order, $Z_g$ and $Z_c$ do not depend on the smearing parameter $\omega$.

\medskip

The final part of our lattice calculation focuses on the computation of Green's functions involving gluino bilinear operators. Although these operators are gauge invariant, their renormalization is extracted from gauge-variant Green's functions. Nevertheless, the corresponding renormalization factors in the $\MSbar$ scheme are expected to be gauge independent. Furthermore, the $1/\epsilon$ poles arising in dimensional regularization (DR) must match the $\log(a)$ divergences in lattice regularization (LR), up to finite terms that depend on the specific lattice discretization.

These finite contributions, particularly those associated with the axial-vector current and the critical gluino mass, can be exploited to determine optimal values for the clover coefficient $c_{\rm SW}$ and the stout smearing parameter $\omega$. One strategy is to enforce the conservation of the axial current on the lattice, which can be achieved by setting the Wilson parameter to $r = 1$, taking $c_{\rm SW} = 1$ in this perturbative order, and appropriately tuning $\omega$. An alternative approach is to choose $c_{\rm SW}$ and $\omega$ such that, with $r = 1$, both the critical gluino mass satisfies $m_\lambda^{\rm crit.} \to 0$ and the axial current renormalization factor approaches unity, $Z_A^{\rm LR,\overline{MS}} \to 1$, thus improving chiral and supersymmetric restoration in the continuum limit.

Below we present the one-loop expressions for the amputated bare Green's functions of the scalar, pseudoscalar, and axial-vector gluino bilinears on the lattice ($\Gamma = \openone,\,\gamma_5,\,\,\gamma_5\,\gamma_\mu$):
\begin{align}
\big\langle \lambda^\alpha(q)\,{\cal O}_\Gamma\,\bar\lambda^\beta(q') \big\rangle^{B,LR}_{\rm amp}
&= (2\pi)^4\,\delta(q-q')\,\frac{1}{4}\,\delta^{\alpha\beta}
\Bigg[
\Gamma
+ \frac{g^2 N_c}{16\pi^2}\Big(
\Gamma\,F_\Gamma + L_\Gamma
\Big)
\Bigg],
\label{eq:generic-green}
\end{align}
where $F_\Gamma$ contains the finite, action-dependent one-loop coefficients.
$L_\Gamma$ denotes the universal logarithmic factor that multiplies $\Gamma$;
furthermore, $L_\Gamma$ contains the universal longitudinal (non-$\Gamma$-multiplicative) term
that appears only for the axial-vector operator, and an integer coefficient that multiplies
the gauge parameter, as shown in Eq.~(\ref{LA}).

Each finite part $F_\Gamma$ is written as:

\begin{equation}
F_\Gamma =
{\rm{c}}_{00}^{\Gamma} 
+ 5.7920 \,\alpha
+ {\rm{c}}_{10}^{\Gamma}\,c_{\rm SW}\,r
+ {\rm{c}}_{20}^{\Gamma}\,c_{\rm SW}^2
+ {\rm{c}}_{11}^{\Gamma}\,c_{\rm SW}\,r\,\omega
+ {\rm{c}}_{01}^{\Gamma}\,\omega
+ {\rm{c}}_{02}^{\Gamma}\,\omega^2,
\label{eq:Fgeneric}
\end{equation}

The universal (action-independent) parts are:
\be
L_S = -(3+\alpha)\,\log(a^2 q^2) \openone, \quad
L_P = -(3+\alpha)\,\log(a^2 q^2) \gamma_5, \quad
L_A = - \alpha\left(1 + \log(a^2 q^2) \right)\gamma_5 \gamma_\mu -2\alpha\,\gamma_5\,\frac{q_\mu \slashed q}{q^2}.
\label{LA}
\ee

\begin{table}[h!]
\centering
\setlength{\tabcolsep}{5pt}
\renewcommand{\arraystretch}{1.2}
\begin{tabular}{llrrrrrrr}
\hline
Operator & Gluon action & ${\rm{c}}_{00}^{\Gamma}$ &  ${\rm{c}}_{10}^{\Gamma}$ & ${\rm{c}}_{20}^{\Gamma}$ & ${\rm{c}}_{01}^{\Gamma}$ & ${\rm{c}}_{11}^{\Gamma}$ & ${\rm{c}}_{02}^{\Gamma}$ \\
\hline
\multirow{3}{*}{$S$ $(\Gamma = 1)$}
 & Wilson   & $0.3080$ &  $19.9736$ &$+0.0675$ &  $17.7450$ & $-55.2027$ & $-79.7022$ \\
 & TLS      & $0.5835$ &  $17.7014$ & $-0.5009$ & $12.3004$ & $-45.7895$ & $-60.0198$ \\
 & Iwasaki  & $0.7409$ &  $13.8034$ &$-1.1734$ &  $5.3684$  & $-31.3677$ & $-34.7285$ \\
\hline
\multirow{3}{*}{$P$ $(\Gamma = \gamma_5)$}
 & Wilson   & $9.9510$ & $0$ & $13.7331$ &  $-58.9422$ & $0$ & $138.5619$ \\
 & TLS      & $8.7101$ & $0$ & $11.9482$ &  $-48.6342$ & $0$ & $108.4866$ \\
 & Iwasaki  & $6.5561$ & $0$ & $9.0153$  &  $-33.5131$ & $0$ & $67.6047$  \\
\hline
\multirow{3}{*}{$A$ $(\Gamma = \gamma_5 \gamma_\mu)$}
 & Wilson   & $-0.8481$ & $4.9934$ & $-3.4164$ & $18.5468$ & $-13.8007$ & $-52.7376$ \\
 & TLS      & $-0.4837$ & $4.4254$ & $- 3.1123$ & $14.7682$ & $-11.4474$  & $-40.7880$ \\
 & Iwasaki  & $+0.0752$ & $3.4508$ & $-2.5472$ & $9.4514$  & $-7.8419$  & $-24.8290$ \\
\hline
\end{tabular}
\caption{Finite coefficients ${\rm c}_i^{\Gamma}$ for the Green's functions $\big\langle \lambda^\alpha(q)\,{\cal O}_\Gamma\,\bar\lambda^\beta(q') \big\rangle^{B,LR}_{\rm amp}$ of the bilinear operators for $\Gamma = \{1, \gamma_5, \gamma_5\gamma_\mu\}$.}
\label{tab:all_coefficients}
\end{table}

In order to extract the renormalization factors from the above results, we impose the following renormalization condition:
\begin{equation}
 \langle \lambda^\alpha \,{\cal{O}}_{\Gamma} \,\bar{\lambda}^{\beta} \rangle^\MSbar_{\text{amp}} = \left(Z_{\lambda}^{LR,\MSbar}\right)^{-1} Z^{LR,\MSbar}_{{\cal{O}}_\Gamma} \langle \lambda^\alpha \,{\cal{O}}_{\Gamma} \,\bar{\lambda}^{\beta} \rangle^{B, LR}_{\rm amp}\Big{|}_{a \to 0},   
\label{GF2Bilinearconf}
\end{equation}
and the renormalization factors are given by the following expressions:

\begin{align}
Z_{{\cal O}_\Gamma}^{LR,\,\overline{\mathrm{MS}}}
&= 1 + \frac{g^2 N_c}{16\pi^2}\,
\Big( {\rm{c}}^{Z_\Gamma}_{00}\;
    + {\rm{c}}^{Z_\Gamma}_{10}\,c_{\rm SW}\,r
    + {\rm{c}}^{Z_\Gamma}_{20}\,c_{\rm SW}^2
    + {\rm{c}}^{Z_\Gamma}_{01}\,\omega
    + {\rm{c}}^{Z_\Gamma}_{11}\,c_{\rm SW}\,r\,\omega
    + {\rm{c}}^{Z_\Gamma}_{02}\,\omega^2
    + \ell^{Z_\Gamma}\,\log(a^2\bar\mu^2)
\Big).
\label{ZOGamma}
\end{align}

\begin{table}[h!]
\centering
\setlength{\tabcolsep}{5pt}
\renewcommand{\arraystretch}{1.15}
\begin{tabular}{llrrrrrrr}
\hline
Operator & Gluon action & ${\rm{c}}^{Z_\Gamma}_{00}$ & ${\rm{c}}^{Z_\Gamma}_{10}$ & ${\rm{c}}^{Z_\Gamma}_{20}$ &  ${\rm{c}}^{Z_\Gamma}_{01}$ & ${\rm{c}}^{Z_\Gamma}_{11}$ & ${\rm{c}}^{Z_\Gamma}_{02}$ & $\ell^{Z_\Gamma}$ \\
\hline
\multirow{3}{*}{$S$ $(\Gamma = 1)$}
 & Wilson   & $-12.9524$ & $-15.4758$ & $+5.5215$ & $+184.2629$ & $+44.0789$ & $-659.9814$ & $+3$ \\
 & TLS      & $-9.6067$  & $-13.6706$ & $+5.4697$ & $+140.2638$ & $+36.4426$ & $-481.3607$ & $+3$ \\
 & Iwasaki  & $-4.8575$  & $-10.6013$ & $+5.0662$ &  $+85.7754$ & $+24.8202$ & $-274.0247$ & $+3$ \\
\hline
\multirow{3}{*}{$P$ $(\Gamma = \gamma_5)$}
 & Wilson   & $-14.5954$ & $+4.4977$ & $-8.1441$ & $+260.9502$ & $-11.1238$ & $-878.2455$ & $+3$ \\
 & TLS      & $-9.7334$  & $+4.0309$ & $-6.9794$ & $+201.1983$ &  $-9.3469$ & $-649.8671$ & $+3$ \\
 & Iwasaki  & $-2.6727$  & $+3.2020$ & $-5.1225$ & $+124.6569$ &  $-6.5475$ & $-376.3579$ & $+3$ \\
\hline
\multirow{3}{*}{$A$ $(\Gamma = \gamma_5\gamma_\mu)$}
 & Wilson   & $-11.7963$ & $-0.4957$ & $+9.0055$ & $+183.4612$ & $+2.6768$ & $-686.9461$ & $0$ \\
 & TLS      & $-8.5396$  & $-0.3945$ & $+8.0811$ & $+137.7960$ & $+2.1005$ & $-500.5925$ & $0$ \\
 & Iwasaki  & $-4.1918$  & $-0.2488$ & $+6.4400$ &  $+81.6924$ & $+1.2944$ & $-283.9241$ & $0$ \\
\hline
\end{tabular}
\caption{Finite coefficients ${\rm{c}}^{Z_\Gamma}_{ij}$ and logarithmic $\ell^{Z_\Gamma}$ coefficients entering 
$Z_{{\cal O}_\Gamma}^{LR,\,\overline{\mathrm{MS}}}$
of bilinear operators with $\Gamma = \{1, \gamma_5, \gamma_5\gamma_\mu \}$ for the three gluon actions (Wilson, TLS, Iwasaki).}
\label{tab:CZ_all_final}
\end{table}

All renormalization factors $Z_{{\cal O}_\Gamma}^{LR,\MSbar}$ are gauge independent, as expected for gauge-invariant composite operators. Among these, only the scalar and pseudoscalar operators exhibit logarithmic divergences. In contrast, the renormalization factor for the axial-vector operator, $Z_A^{LR,\MSbar}$, is finite at one-loop order and does not depend on the renormalization scale. This behavior reflects the (partial) conservation of the corresponding current and is consistent with Ward identities in the continuum theory.

\begin{figure}[ht!]
\centering
\includegraphics[width=0.85\textwidth]{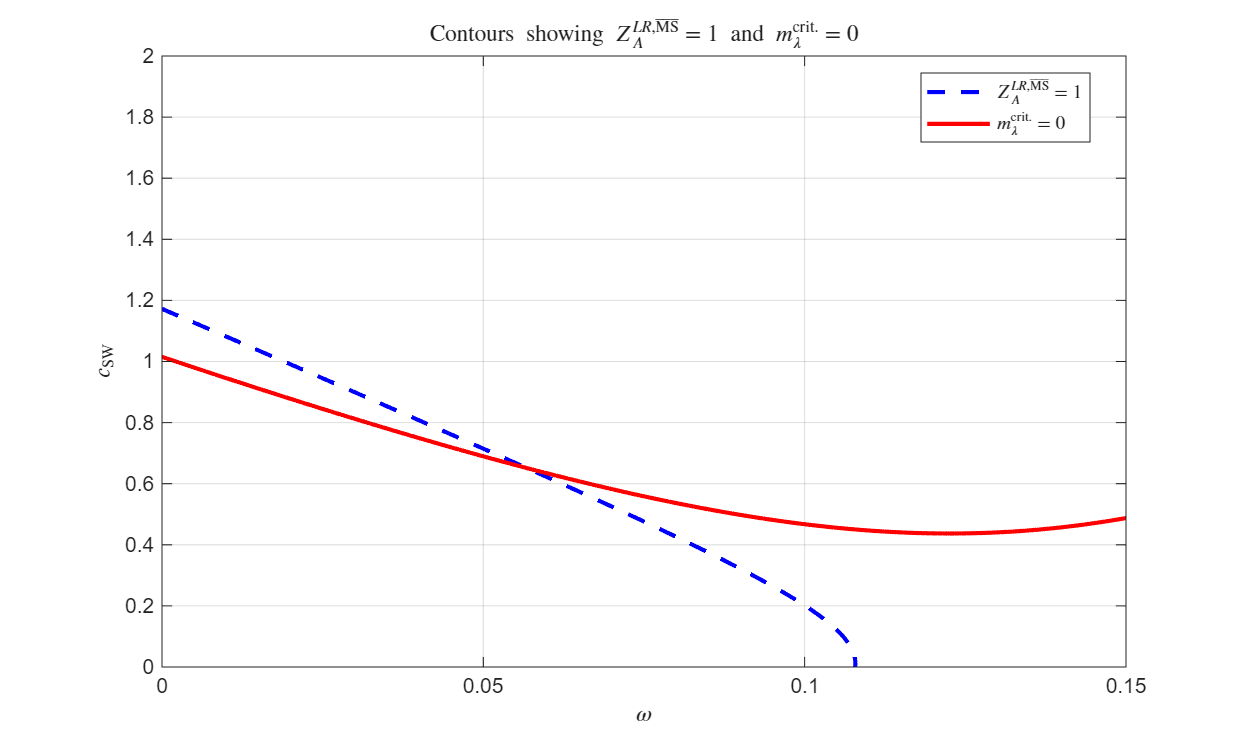}
\caption{Contours of the two conditions in the $(\omega, c_{\mathrm{SW}})$ plane:
$Z_A^{LR,\overline{\rm{MS}}}=1$ (blue, dashed) and 
$m_\lambda^{\mathrm{crit.}}=0$ (red, solid) for the Wilson action.
Their intersection determines the values of $c_{\mathrm{SW}}$ and $\omega$ 
for which both conditions are simultaneously satisfied.}
\label{fig:contours}
\end{figure}

An interesting choice of parameters, as shown in Fig.~\ref{fig:contours}, is to fix $r=1$ but select $c_{\mathrm{SW}}$ and $\omega$ such that both targets for the 
critical mass, $m_\lambda^{\mathrm{crit.}}=0$, and for the axial renormalization factor, $Z_A^{LR,\overline{\mathrm{MS}}}=1$, are simultaneously met. This leads to 
a unique solution which, in the case of the Wilson gluon action, reads: $c_{\mathrm{SW}} \approx 0.6512, \,\omega \approx 0.0568$\,.

\section{Conclusions -- Future plans}
\label{future}

In this work, we present a perturbative investigation of renormalization factors in lattice $\mathcal{N}=1$ supersymmetric Yang–Mills (SYM) theory, employing a fermion action with stout-smeared links and  Symanzik-improved gluon actions. We perform one-loop calculations of two- and three-point Green’s functions within the lattice regularization, allowing us to derive explicit expressions for the renormalization factors of the gluon, gluino and ghost fields, the gauge parameter, the coupling constant and the critical gluino mass. Further, we compute the renormalization factors for a complete set of gluino bilinear operators. These expressions are presented in fully general form, retaining the dependence on the number of colors $N_c$, the gauge parameter $\alpha$, the clover coefficient $c_{\rm SW}$, and the stout smearing parameter $\omega$, while results for the Wilson parameter are given for $r = \pm 1$. 

An outcome of this study is that the axial-vector renormalization constant, $Z_A$, can be conserved at one loop for a particular choice of the stout-smearing parameter. 

Furthermore, we can make also an appropriate choice for the clover coefficient, leading to a vanishing critical mass for the gluino; i.e., we determine pairs $(c_{\rm SW},\omega)$ (with $r=1$) such that simultaneously $m_\lambda^{\mathrm{crit.}}\to 0$
and $Z_A^{\rm LR,\overline{MS}}\to 1$, cf. Eqs.~(\ref{mcrit}),~(\ref{ZOGamma}).

Our results confirm that stout smearing can significantly reduce lattice discretization effects, allowing for a more stable computation of renormalization factors in lattice SYM. The use of stout-smeared links improves perturbative convergence and enables appropriate fine-tuning. Future work includes the renormalization of the gluino–glue operator and the determination of the mixing matrix of the Noether supercurrent in the presence of stout smearing. These calculations are important for a better control of supersymmetry on the lattice and providing input to nonperturbative studies.

\appendix
\section{The path integral over the gluino field}
\label{majoranaS}

To elucidate the Majorana nature of the gluino field within the functional integral, and the way to properly address it in the calculation of Feynman diagrams, we first reformulate the SYM action~\cite{Donini:1997hh}, expressing it exclusively in terms of $\lambda$, rather than $\bar \lambda$. Applying the Majorana condition ($(\bar \lambda^{\alpha})^T= C\lambda^{\alpha}$), the part of the action which contains gluino fields has the general form:
\begin{equation}
S_{\text{gluino}}= \bar \lambda D \lambda  = \lambda^T M \lambda, 
\end{equation}
where, making explicit the spacetime ($x,\ y$), color adjoint ($\alpha,\ \beta$) and spinorial ($i,\ j$) indices of the matrix $D$, it reads: $(D_{xy})^{\alpha\,\beta}_{i\,j}$\,. The antisymmetric matrix $M$ equals: $M \equiv C^T D$.
Therefore, the path integral reads:
\begin{equation}
    Z[J] = \int \mathcal{D} X\,  e^{-S_{X}} \int \mathcal{D} \lambda \, e^{- \lambda^T M \lambda  - J \lambda } \, ,
    \label{PartFun}
\end{equation}
where $J$ is an external Grassmann source, $X$ stands for the other fields apart from the gluino fields, and $S_{X}$ denotes the part of the action devoid of gluinos. Note that we do not assume that $J$ is Majorana spinor. In order to integrate out the gluino field, we implement the following standard change of variables:
\begin{equation}
    \la' \equiv \la - \frac{1}{2}  \, M^{-1}\, J^T
    \label{laprime}
\end{equation}
This leads to:

\begin{equation}
    Z[J] = \int \mathcal{D} X\,  e^{-S_X} \int \mathcal{D} \lambda' \, e^{- \lambda'^T M \lambda'  - \frac{1}{4} J M^{-1} J^T} = \, \int \mathcal{D} X\,  e^{-S_X} Pf[2M] \, e^{- \frac{1}{4} J M^{-1} J^T} \equiv \int_{\!X} \, e^{- \frac{1}{4} J M^{-1} J^T},
   \label{ZJeq}
\end{equation}
where $Pf[2M]$ is the Pfaffian of the antisymmetric matrix $2M$. The factor of 2 appearing in the Pfaffian is inconsequential, and it will be dropped from this point on. In case one is interested only in Green's functions without external gluinos (so that one can set $J=0$ from the start), the exponential in Eq.~(\ref{ZJeq}) becomes trivial and the only remnant of gluinos is the Pfaffian; in those cases, the only effect of the gluinos' Majorana nature is the well-known factor of 1/2 for every closed gluino loop, due to the fact that $Pf[M] = \rm{det}[M]^{1/2}$.  

In order to compute Green's functions with two external gluinos, for example $\la(x) \bar \la(y)$, we have to consider the following second derivative with respect to the external source\footnote{In what follows, $M^{-1}_{x y}$ is shorthand for $(M^{-1})_{x y}$; similarly for the matrix $D^{-1}$.}:
\begin{align}
    \langle\la_i^\alpha(x) \bar\la_j^\beta(y) \rangle=  \langle\la_i^\alpha(x) \la_{j'}^\beta(y) \mathcal{C}_{j\,j'}  \rangle &= \frac{1}{Z[0]}\bigg(-\frac{\delta}{\delta J_i^\alpha(x)}\bigg) \bigg(-\frac{\delta}{\delta J_{j'}^\beta(y)}\bigg) \, Z[J]\Bigg{|}_{J=0} \,\mathcal{C}_{j\,j'} \nonumber\\[0.6em]
    &= \frac{1}{2}\,\frac{1}{Z[0]}\int_{\!X}  \left(M^{-1}_{x y}\right)_{i\,j'}^{\alpha\,\beta} \, \mathcal{C}_{j\,j'} = \frac{1}{2}\,\frac{1}{Z[0]}\int_{\!X} \left(D^{-1}_{x y}\right)_{i\,j}^{\alpha\,\beta}.
\end{align}
Therefore, an overall factor of $1/2$ (not stemming from the Pfaffian) is needed for the gluino propagator; this factor is again due to the Majorana nature of the gluino.

Gluon fields are contained in the matrices $M^{-1}$ and $D^{-1}$, and can be extracted via a series expansion in $g$; thus, one gluon emerges by calculating the quantity $g \frac{\partial}{\partial g}(M^{-1})\Bigr|_{g=0}$:
\begin{align}
    & g \frac{\partial}{\partial g} (M^{-1})\Bigg|_{g=0} = -M^{-1} \bigg( g \frac{\partial M}{\partial g} \bigg) M^{-1} \Bigg|_{g=0}
\end{align}
where $g \frac{\partial M}{\partial g}$ is the normal vertex with two gluinos and one gluon. Similarly, extraction of two gluons follows from:
\begin{align}
     \frac{1}{2} g^2 \frac{\partial^2}{\partial g^2} (M^{-1})\Bigg|_{g=0} & = - \frac{1}{2} \, g^2 \, \frac{\partial}{\partial g} \bigg( M^{-1} \frac{\partial M}{\partial g} M^{-1} \bigg)\Bigg|_{g=0}  \nonumber \\
    & = g^2 M^{-1} \bigg(\frac{\partial M}{\partial g} \bigg) M^{-1} \bigg( \frac{\partial M}{\partial g} \bigg) M^{-1} \Bigg|_{g=0} - \frac{1}{2} \, g^2 M^{-1} \, \frac{\partial^2 M}{\partial g^2} \, M^{-1}
\end{align}
The term with $\frac{\partial^2 M}{\partial g^2}$ appears only on the lattice.

We turn now to more general Green's functions involving gluinos. As a prototype case, let us consider the Green's function of a gluino bilinear operator with two external gluino fields, $\langle \lambda^\alpha \,{\cal{O}}_{\Gamma} \,\bar{\lambda}^{\beta} \rangle $. This can be derived from the functional integral by taking four functional derivatives, as follows:

\bea
\sum_{z}\,\langle \lambda_i^\alpha(x)\, \frac{1}{2} \bar\lambda^\gamma_k(z)\, \Gamma_{k\, m}\, \lambda^\gamma_m(z)\, \bar\lambda_j^\beta(y) \rangle &=& \sum_{z}\,\langle \lambda_i^\alpha(x)\, \frac{1}{2} \lambda^\gamma_{k'}(z) \,\mathcal{C}_{k\,k'}\, \Gamma_{k\, m}\, \lambda^\gamma_m(z)\, \lambda_{j'}^\beta(y) \, \mathcal{C}_{j\,j'} \rangle
\nonumber \\[0.6em]
&=& \frac{1}{Z[0]}\sum_{z}  \frac{1}{2} 
\frac{\delta}{\delta J_i^\alpha(x)}\,
\frac{\delta}{\delta J_{k'}^\gamma(z)}\,\mathcal{C}_{k\,k'}\, \Gamma_{k\,m}\, \frac{\delta}{\delta J_m^\gamma(z)}\, \frac{\delta}{\delta J_{j'}^\beta(y)}\,\mathcal{C}_{j\,j'}\,Z[J]\Bigg{|}_{J=0} 
\eea

The first derivative, $\delta/\delta J_{j'}^{\beta}(y)$, acts on $Z[J]$ and gives:
\bea
\frac{\delta}{\delta J_{j'}^{\beta}(y)}\,Z[J]
&=&
\int_{\!X}\left(
-\frac{1}{4}\,\left(M^{-1}_{y y'}\right)^{\beta \delta}_{j' j''}\,J^{\delta}_{j''}(y')
+\frac{1}{4}\,J^{\delta}_{j''}(y')\,\left(M^{-1}_{y' y}\right)^{\delta \beta}_{j'' j'}
\right) \, e^{- \frac{1}{4} J M^{-1} J^T}
\nonumber\\[0.6em]
&=&
\int_{\!X}-\frac{1}{2}\,\left(M^{-1}_{y y'}\right)^{\beta \delta}_{j' j''}\,J^{\delta}_{j''}(y')\, e^{- \frac{1}{4} J M^{-1} J^T},
\qquad \text{since $M$ (hence $M^{-1}$) is antisymmetric.}
\eea

Continuing, the second derivative $\delta/\delta J_m^{\gamma}(z)$ acts on the result of the first one, and gives
\bea
\frac{\delta}{\delta J_{m}^{\gamma}(z)}\,
\frac{\delta}{\delta J_{j'}^{\beta}(y)}\,Z[J]
&=&
\int_{\!X}\left(-\frac{1}{2}\,\left(M^{-1}_{y z}\right)^{\beta \gamma}_{j' m}
-\frac{1}{4}\,
\left(M^{-1}_{y y'}\right)^{\beta \delta}_{j'j''}\,J^{\delta}_{j''}(y')\,
\left(M^{-1}_{z z'}\right)^{\gamma \delta'}_{m j'''}\,J^{\delta'}_{j'''}(z')\right) \, e^{- \frac{1}{4} J M^{-1} J^T}.
\label{eq:second-deriv}
\eea
Here the first term comes from differentiating the explicit source $J(y')$, while the second term comes from differentiating the exponential.

We now apply~\eqref{eq:second-deriv} to the Green's function and act with the remaining two derivatives:
\bea
&&\hspace{-1,78cm}
\frac{1}{Z[0]}\int_{\!X}\sum_{z}\,
\frac{1}{2}\,
\mathcal{C}_{k k'}\,\Gamma_{k m}\,\mathcal{C}_{j j'}\,
\frac{\delta}{\delta J_i^{\alpha}(x)}\,
\frac{\delta}{\delta J_{k'}^{\gamma}(z)}\,
\Bigg[
-\frac{1}{2}\,\left(M^{-1}_{y z}\right)^{\beta \gamma}_{j' m}\nonumber\\[0.6em]
&&\hspace{5.2cm}-\frac{1}{4}\,
\left(M^{-1}_{y y'}\right)^{\beta \delta}_{j'j''}\,J^{\delta}_{j''}(y')\,
\left(M^{-1}_{z z'}\right)^{\gamma \delta'}_{m j'''}\,J^{\delta'}_{j'''}(z')
\Bigg]\, e^{- \frac{1}{4} J M^{-1} J^T}
\Bigg|_{J=0}.
\label{eq:after-second}
\eea

\medskip
\noindent
In taking the third derivative, $\delta/\delta J_{k'}^{\gamma}(z)$, we need only keep terms linear in $J$, since any higher order terms will vanish when we set $J=0$ in the end.
The above expression becomes:
\bea
&&\hspace{-1cm}
\frac{1}{Z[0]}\int_{\!X}\sum_{z}\,
\frac{1}{2}\,
\mathcal{C}_{k k'}\,\Gamma_{k m}\,\mathcal{C}_{j j'}\,
\frac{\delta}{\delta J_i^{\alpha}(x)}\,\Bigg[
\frac{1}{4}\,\left(M^{-1}_{y z}\right)^{\beta \gamma}_{j' m}\,
\left(M^{-1}_{z x'}\right)^{\gamma \alpha'}_{k' i'}\,J^{\alpha'}_{i'}(x')
\nonumber\\[0.6em]
&&
-\frac{1}{4}\,\left(M^{-1}_{y z}\right)^{\beta \gamma}_{j' k'}\,
\left(M^{-1}_{z x'}\right)^{\gamma \alpha'}_{m i'}\,J^{\alpha'}_{i'}(x')
+\frac{1}{4}\,
\left(M^{-1}_{z z}\right)^{\gamma \gamma}_{m k'}\,
\left(M^{-1}_{y x'}\right)^{\beta \alpha'}_{j' i'}\,J^{\alpha'}_{i'}(x')
\;+\;\mathcal{O}(J^{3})\Bigg]\, e^{- \frac{1}{4} J M^{-1} J^T}
\Bigg|_{J=0}.
\label{eq:third-deriv}
\eea

The term proportional to $\left(M^{-1}_{zz}\right)^{\gamma\gamma}_{m k'}$ corresponds to diagrams in which the $\lambda$ and $\bar\lambda$ emanating from the bilinear operator form a closed fermion line. A subset of these diagrams (including the $\mathcal{O}(g^2)$ case) will be completely disconnected, and they will contain the vacuum expectation value of $\mathcal{O}_\Gamma$\,.

Therefore, if one is interested in the connected Green's function, one can work directly with the vacuum-subtracted operator $\mathcal{O}_\Gamma\;\longrightarrow\; \mathcal{O}_\Gamma-\langle \mathcal{O}_\Gamma\rangle$.

In this case the purely disconnected contributions drop out. 

Finally, the last derivative $\delta/\delta J_i^{a}(x)$ removes the remaining source, so that, keeping only the connected contributions, the Green's function becomes:
\bea
&&\hspace{-1cm}\frac{1}{Z[0]}\,
\sum_{z}\,
\frac{1}{2}\,
\mathcal{C}_{k k'}\,\Gamma_{k m}\,\mathcal{C}_{j j'}\,
\frac{\delta}{\delta J_i^{\alpha}(x)}\,
\frac{\delta}{\delta J_{k'}^{\gamma}(z)}\,
\frac{\delta}{\delta J_{m}^{\gamma}(z)}\,
\frac{\delta}{\delta J_{j'}^{\beta}(y)}\,Z[J]\Bigg|_{J=0}
\nonumber\\[0.8em]
&&\hspace{-0.5cm}=
\frac{1}{Z[0]}\,\int_{\!X}\sum_{z}\,
\frac{1}{2}\,
\mathcal{C}_{k k'}\,\Gamma_{k m}\,\mathcal{C}_{j j'}\,
\frac{1}{4}\Bigg[
\left(M^{-1}_{x z}\right)^{\alpha \gamma}_{i k'}\,
\left(M^{-1}_{z y}\right)^{\gamma \beta}_{m j'}
-\left(M^{-1}_{x z}\right)^{\alpha \gamma}_{i m}\,
\left(M^{-1}_{z y}\right)^{\gamma \beta}_{k' j'}
+\left(M^{-1}_{z z}\right)^{\gamma \gamma}_{m k'}\,
\left(M^{-1}_{y x}\right)^{\beta \alpha}_{j' i}\Bigg].
\label{eq:four-deriv-result}
\eea
One may then use $M^{-1}=-D^{-1}\mathcal{C}^{-1}$ to convert~\eqref{eq:four-deriv-result} into the
equivalent expression in terms of $D^{-1}$ and, for $S,P,A$, combine the first two contractions into a single
$D^{-1}\Gamma D^{-1}$ form.

\bea
&&
\sum_{z}\,
\frac{1}{8}\,
\Big(
\,\left(M^{-1}_{x z}\right)^{\alpha \gamma}_{i k'}\,
\mathcal{C}_{k k'}\,\Gamma_{k m}\,
\left(M^{-1}_{z y}\right)^{\gamma \beta}_{m j'}\,
\mathcal{C}_{j j'}
-\,\left(M^{-1}_{x z}\right)^{\alpha \gamma}_{i m}\,
\mathcal{C}_{k k'}\,\Gamma_{k m}\,
\left(M^{-1}_{z y}\right)^{\gamma \beta}_{k' j'}\,
\mathcal{C}_{j j'}
\Big)
\nonumber\\[0.8em]
&=&
\sum_{z}\,
\frac{1}{8}\,
\Big(
\left(D^{-1}_{x z}\right)^{\alpha \gamma}_{i k}\,
\Gamma_{k m}\,
\left(D^{-1}_{z y}\right)^{\gamma \beta}_{m j}
+\,\left(D^{-1}_{x z}\right)^{\alpha \gamma}_{i m'}\,
\Gamma_{m' k'}\,
\left(D^{-1}_{z y}\right)^{\gamma \beta}_{k' j}
\Big)
\quad \text{(use $M^{-1}=-D^{-1}\mathcal{C}^{-1}$)}
\nonumber\\[0.8em]
&=&
\frac{1}{2}\, \sum_{z}\,\frac{1}{2}\,
\left(D^{-1}_{x z}\right)^{\alpha \gamma}_{i k}\,
\Gamma_{k m}\,
\left(D^{-1}_{z y}\right)^{\gamma \beta}_{m j}
\eea

Thus, the final form of the Green's function $\langle \lambda^\alpha \,{\cal{O}}_{\Gamma} \,\bar{\lambda}^{\beta} \rangle $ reads:

\bea
&&\frac{1}{2}\,\frac{1}{Z[0]}\,\int_{\!X} \left[\sum_{z}\,\frac{1}{2}\,
\left(D^{-1}_{x z}\right)^{\alpha \gamma}_{i k}\,
\Gamma_{k m}\,
\left(D^{-1}_{z y}\right)^{\gamma \beta}_{m j}
-\frac{1}{4}\left(D^{-1}_{x y}\right)^{\alpha\beta}_{i j} 
\left( \sum_z\,\Gamma_{km}\left(D^{-1}_{zz}\right)^{\gamma\gamma}_{mk}\right) \right]
\eea

Consequently, for Majorana fermions, the two terms (connected and disconnected) in the  resulting Green's function differ from those in the Dirac case by overall factors of $1/2$ and $1/4$, respectively; once again, these factors are not the standard ones which stem from the Pfaffian.

\begin{acknowledgments}
 The project is implemented under the programme of social cohesion``THALIA 2021-2027'' co-funded by the European Union through the Research and Innovation Foundation (RIF). 
\end{acknowledgments}

\end{document}